\documentclass[a4paper,twoside]{article}
\newcommand{\affil}[1]{$^{\rm #1}$}
\usepackage[authoryear]{natbib}
\bibpunct{(}{)}{;}{a}{}{,}
\usepackage{graphicx}
\date{} 
\newcommand{\um}{$\mu$m}
\newcommand{\apj}{ApJ}

\newcommand{\apjs}{ApJS}
\newcommand{\aj}{AJ}
\newcommand{\nat}{Nature}
\newcommand{\aap}{A\&A}
\newcommand{\aas}{A\&ASS}
\newcommand{\pasp}{PASP}

\newcommand{\mnras}{MNRAS}%

\title{\large\bf\flushleft A Review of AGB Mass Loss Imaging Techniques}
\author{\parbox{\textwidth}{\flushleft
\vspace{-0.5cm}
{\it M. Marengo\affil{A,B}}\\
\vspace{0.4cm}
{\small \affil{A}\,Harvard-Smithsonian Center for Astrophysics,
  Cambridge, MA 02138, USA}\\
{\small \affil{B}\,Email: mmarengo@cfa.harvard.edu}}}
\begin{document}
\twocolumn[
\begin{changemargin}{.8cm}{.5cm}
\begin{minipage}{.9\textwidth}
\vspace{-1cm}
\maketitle
%
%
\small{\bf Abstract:}

Circumstellar imaging, across the electromagnetic spectrum, allows to
derive fundamental diagnostics for the physics of mass loss in the AGB
phase. I review the current status of the field, with particular
emphasis on the techniques that provide the strongest constraints for
mass loss modeling efforts.


\medskip
\medskip
\end{minipage}
\end{changemargin}
]
\small

\section{Introduction}
\label{sec-intro}

Asymptotic Giant Branch (AGB) stars are the ancient alchemists dream:
complex nucleosynthesis and mixing processes in their nuclear furnaces
are the source of a variety of heavy nuclei that are ultimately
released into the interstellar medium by strong stellar winds. With a
cumulated rate of $\sim 0.55$~M$_\odot$~yr$^{-1}$ (75\% of the total
mass loss in the Galaxy), AGB stars are among the principal culprits
in the heavy elements pollution of the Galaxy \citep{Sedlmayr1994}.

Despite their importance for stellar evolution and the chemical
evolution of the interstellar medium, the processes responsible for
the AGB winds are not fully understood. While there is a clear link
between mass loss and stellar pulsations (AGB stars tend to be Long
Period Variables), the details of this relation are not completely
known. The basic mechanism appears to be radiation pressure on dust
grains forming in their extended atmospheres \citep{Salpeter1974}, but
a self consistent theory predicting AGB mass loss rates is
still lacking. While models of dust driven winds in C-rich (C/O$>$1)
AGB atmospheres are now capable of passing observational tests (see
e.g. \citealt{Gautschy-Loidl2004, Nowotny2005}), the same does not
hold true for silicate dust in an O-rich environment
\citep{Woitke2006, Hofner2007, Hofner2008}. A comprehensive set of
prescriptions for mass loss rates, dust production and yields as a
function of stellar parameters (including metallicity) and time are
still missing and highly desirable \citep{vanLoon2008}.

Even the basic mass loss parameters (the mass loss rate, the total
mass released by the star during the AGB phase, and the dust to gas
mass ratio) are poorly determined by current observational techniques
\citep{Ramstedt2008, vanLoon2000}. While spectral observations are
crucial to determine the kinematic and composition of the AGB wind,
direct imaging of the circumstellar envelopes provide unique
spatial information.  Images at different wavelengths allow probing
the spatial structure, location and temporal variation of the mass
loss ``engine'', and a ``fossil record'' of the AGB mass loss.


While this is not aimed to be a comprehensive review, within the next
sections I will summarize recent results for current imaging
techniques designed to probe the different aspects of the AGB
outflows. Radio and sub-mm maps are sensitive to the molecular gas
components, and are essential to estimate the total mass loss rate of
the stars and its detailed chemistry
(Section~\ref{sec-radio}). Optical scattered light and polarimetry are
instead an important tool to map the dust content of the circumstellar
envelope at large distances from the star, thus providing a
historical record of the mass loss rate over tens of thousand of years
(Section~\ref{sec-scattered}). The infrared domain is sensitive to the
thermal emission of the circumstellar dust. As such it is the
preferred domain to probe the inner regions of the envelopes, where
dust formation and growth processes are active
(Section~\ref{sec-ir}). High energy phenomena like jets and shocks can
arise when the AGB star is part of a close binary system. Studying
these structures is important to understand the physics of mass
transfer and the origin of asymmetries in the AGB wind
(Section~\ref{sec-xrays}). Section~\ref{sec-star} instead focuses on
the imaging of the star itself, in its role as engine of the mass loss
processes. In Section~\ref{sec-future} I will conclude by exploring
the opportunities that will be offered for AGB circumstellar envelopes
imaging with the instrumentation available within the next decade.


\section{The Molecular Envelopes}
\label{sec-radio}

Most of the gas phase in AGB circumstellar envelopes is in molecular
form, and can be mapped with radio and millimeter-wave
observations. Since the first detection of OH masers around evolved
stars \citep{Wilson1968}, and CO in the envelope of the prototype
carbon star IRC$+$10216 \citep{Solomon1971}, a large number of
molecules have been detected \citep{Olofsson2006}. Mapping the
distribution of these molecules with single dish and interferometric
observations allows to study the geometry of the gas component in the
AGB molecular outflows.

Thermally excited molecular transitions (traditionally CO and HCN, but
more recently also rotational lines from SiO) are observed in AGB
stars of all chemical types, and are used as reliable indicators of
mass loss \citep{Loup1993, Gonzalez-Delgado2003} and to study the
general symmetry of the molecular envelope \citep{Neri1998,
  Olofsson1996, Olofsson2000, Lindqvist1999, Fong2006}. High angular
resolution imaging of those molecules has recently provided the
measurement of velocity gradients, anisotropies and chemical
fractionation in the molecular envelopes (see
e.g. \citealt{Muller2008, Dinh-V-Trung2008, Schoier2006a}), as well as
opened a new avenue to study non-LTE chemistry and grain adsorption
processes \citep{Schoier2006b}. Multi-epoch imaging of SiO, OH, H$_2$O
and other masers (see e.g. the ``movie'' of SiO maser emission
detected around the Mira variable TX Cam by \citealt{Diamond2003}),
allows to probe with sub-mas resolution the local gas infall, outflow
and non-radial motion superposed to the dominant expansion of the
circumstellar envelope. The polarization of SiO and H$_2$O maser lines
has also provided indications of stellar and circumstellar magnetic
fields (see e.g. \citealt{Kemball1997, Vlemmings2001, Vlemmings2002}),
and allowed to observe the emergence of collimated jets at the end of
the AGB phase \citep{Vlemmings2006}.

Observations of isotopologues of several molecules (for example
$^{12}$CO and $^{13}$CO, \citealt{Milam2009} and references therein)
allow to study the effects of stellar nucleosynthesis and transport
processes \citep{Soria-Ruiz2005}. These observations will become very
important when the Atacama Large Millimeter Array (ALMA) will be
operational starting 2010. The high resolution ($\sim 0.02$~arcsec)
and sensitivity of ALMA, between 0.3 and 9.6~mm, will allow to
spatially resolve the crucial isotopic and chemical gradients within
the AGB circumstellar shells, reflecting the chronology of otherwise
inaccessible stellar nuclear processes.

The molecular envelope of AGB and post-AGB stars can also be imaged in
H$_2$ line emission with narrow band filters at wavelengths ranging
from the UV to the IR. The long tail of Mira ($o$~Cet), tracing 30~000
years of mass loss history and interactions with the interstellar
medium, has been detected in far-UV H$_2$ line with GALEX
\citep{Martin2007}.  Spectral H$_2$ imaging in the infrared has been
applied mostly to map the morphology of post-AGB stars and the AGB
remnant halo of Planetary Nebul\ae{} \citep{Hrivnak2008, Hrivnak2006,
  Sahai1998, Kastner2001, Cox2003, Davis2005, VandeSteene2008}.  The
novel technique of Integral Field Spectroscopy has been applied for
these evolved objects to obtain higher spatial and spectral resolution
in molecular hydrogen and other lines \citep{Tsamis2008, Sandin2008,
  Santander-Garcia2008, Santander-Garcia2007, Sharp2006, Lowe2006,
  Monreal-Ibero2005}.

While the gas in the AGB circumstellar envelope is mostly molecular,
recent observations have been able to detect atomic lines. The 21~cm
H{\sc i} neutral atomic hydrogen line, in particular, allows to probe
AGB outflow at large distance from the star \citep{Gerard2006,
  LeBertre2004, Gardan2006, Matthews2007}. Notable is the case of the
detection of the H{\sc i} counterpart of the $o$~Cet tail
\citep{Matthews2008}. \citet{Reid2007} have imaged at VLA the radio
continuum emission from H$^-$ free-free interactions in the stellar
photosphere of 3 AGB stars, finding asymmetries in two of them (R~Leo
and W~Hya, while $o$~Cet appeared to be spherically symmetric). These
observations, if they can be obtained as time series for a larger
sample of stars, will play an important role in understanding the
physics at the interface between the stellar atmosphere and the
circumstellar envelope, to better characterize the role of stellar
pulsations in triggering the mass loss.


\section{Fossil Dust Shells in Scattered Light}
\label{sec-scattered}

Circumstellar envelopes keep the historical record of the mass loss
rate variations during the AGB phase. Far from the star the released
material is not directly excited by the stellar radiation, but its
dust component can still be observed at optical wavelengths in
scattered light. \citet{Mauron1999, Mauron2000} first detected
concentric shell structures around IRC$+$10216, hinting to alternating
phases of high and low mass loss in the last $\sim $8~000~yr with a
cadence of $\sim 200$--800~yr. Large aperture ground based telescopes
like the VLT are allowing to probe with this technique the geometry of the
envelopes and the mass loss history for an increasing number of
sources at different evolutionary stage along the AGB
\citep{Mauron2006}.

Imaging polarimetry in the optical and in the near-IR enhance the
detection of scattered light by dust grains in faint circumstellar
structures by suppressing the light from the central star. Low
resolution polarimetric maps of circumstellar envelopes
have been obtained for more than 15 years \citep{Kastner1992,
  Kastner1994}. The availability of polarimeters on the HST telescope
and at large aperture ground based telescopes equipped with Adaptive
Optics (AO) have recently allowed probing the asymmetry of dusty
shells at sub-arcsec angular scale for sources at different
evolutionary stage (see e.g. \citealt{Gledhill2001, Gledhill2005,
  Ueta2007, Murakawa2005, Murakawa2008}).


\section{Dust Thermal Radiation}
\label{sec-ir}

Thermal infrared is the traditional wavelength range to study the
dust component of AGB circumstellar envelopes. It is no surprise that
the first astronomical observations with infrared bolometers were cool
evolved stars \citep{Low1964, Neugebauer1965}, soon recognized to be
hosting circumstellar envelopes of silicate and carbonaceous dust
\citep{Wolf1969}. The first ``image'' of an AGB circumstellar envelope
in the thermal infrared was the determination of the angular diameter
of IRC$+$10216 by means of a lunar occultation
\citep{Toombs1972}. While lunar occultations are still used today to
image dust shells around AGB stars (see e.g. \citealt{Harvey2007}),
the current state of the art in high resolution thermal infrared
imaging of AGB circumstellar envelopes is direct imaging with
AO systems and interferometric techniques.

\subsection{Direct Imaging}
\label{ssec-imaging}

Thermal infrared astronomy literally took off with the launch of the
InfraRed Astronomical Satellite (IRAS) in 1983. A careful analysis of
IRAS data found 15 sources associated to evolved stars that were
showing extended emission at 60 and 100~\um{} \citep{Hacking1985}. A
reprocessing of the IRAS data to create higher resolution maps using
Pyramid Maximum Entropy (PME) image reconstruction technique (HIRAS,
\citealt{Bontekoe1994}), allowed the detection of cold detached shells
around three of these stars: the carbon stars U~Hya \citep{Waters1994}
and U~Ant \citep{Izumiura1997} and the O-rich star R~Hya
\citep{Hashimoto1998}. A similar analysis with the Infrared Space
Observatory (ISO) photometer (ISOPHOT) revealed a dust shell around
the J-type carbon star Y~CVn \citep{Izumiura1996}. The age of the shells
($\sim 10^3$--$10^4$~yr, 10 to 100 times longer than the shells
detected in scattered light around IRC$+$10216 and other evolved AGB
stars) suggests that these phases of increased mass loss may be
related to the thermal pulses in the TP-AGB phase \citep{Willems1988,
  Zijlstra1992}. A search for more of these detached shells is
currently being made using the MIPS instrument on-board the Spitzer
Space Telescope \citep{Ueta2006, Ueta2008a} and the FIS instrument on
the AKARI Infrared Astronomy Satellite \citep{Ueta2008b}.

The development of array cameras capable of obtaining background
limited images in the N ($\sim 10$~\um) and Q ($\sim 20$~\um)
atmospheric windows allowed the observation of AGB circumstellar dust
from the ground. The first observational campaigns focused on post-AGB
stars, because of their high infrared excess and large size
\citep{Dayal1998, Meixner1999, Ueta2001, Kwok2002}. Early observations
of stars still on the AGB \citep{Busso1996, Sudol1999, Marengo1999,
  Marengo2001, Jura2002a, Jura2002b, Lagadec2005} could only
partially resolve the circumstellar envelope. Their main limiting
factor was in the stability of the instrumental Point Spread Function
(PSF), required for the removal of the bright stellar source to
isolate the fainter circumstellar emission. This limitation was
resolved with the introduction of AO systems.

The main advantage of AO is that it guarantees a very stable PSF by
compensating for the seeing and for distortions in the optical
system. In the thermal infrared this yields diffraction limited
images with a very high Strehl ratio (98\% or better) which do not
significantly change while moving from one target to another and to
a reference star. Using AO it is possible to isolate small scale
structures and detect deviations from spherical symmetry with high
reliability even in sources that are only marginally extended with
respect to the reference PSF \citep{Biller2005, Biller2006, Leao2006,
  Lagadec2007}.

\subsection{Interferometric Techniques}
\label{ssec-interferometers}

While radio interferometry has been used for many decades to acquire
astronomical images, infrared interferometry is just now reaching
maturity. With interferometry it is possible to record both phases and
amplitudes of incident light by combining the beams from two or more
apertures. With long baseline interferometry the distance between the
apertures can be of hundreds of meters, allowing to reach milli-arcsec
angular resolution. For nearby AGB stars this is sufficient to resolve
details of the stellar disk and map with great accuracy the
anisotropies and asymmetries in the dust forming
region. Interferometry techniques can also be applied to single
aperture telescopes by subdividing the pupil into sub-apertures,
either with a physical mask (aperture masking), or by imaging seeing
speckles (speckle imaging).

The first interferometric observations of dusty circumstellar
envelopes in the thermal infrared became possible with the development
of the Infrared Spatial Interferometer (ISI, \citealt{Danchi1990,
  Hale2000}). ISI is unique among the infrared interferometers because
it combines the light from its individual apertures (originally two
truck-mounted 1.65~m movable telescopes) using a heterodyne detector
with $^{13}$CO$_2$ lasers and local oscillators. ISI operates at
11~\um{} with baselines ranging from 4 to 70~m (providing a maximum
angular resolution of $\sim 20$~mas), and since 1991
\citep{Bester1991} it has been used to derive the diameters and radial
profiles for a large number of AGB circumstellar envelopes. Recently,
the interferometer has been equipped with a third telescope, allowing
it to use closure phases as a mean to record with great accuracy the
symmetry of the observed sources \citep{Weiner2006a, Tatebe2006,
  Chandler2007a, Chandler2007b}.

The MIDI instrument \citep{Leinert2003} operates in the N and Q
infrared bands by combining the light from two of the VLTI telescopes
(either the large 8~m UTs or the 1.8~m ATs) with baselines ranging
from 8 to 200~m. This provides an angular resolution of up to $\sim
20$~mas within a field of view of up to $\sim 1$~arcsec radius. The
limiting sensitivity of the instrument is $\sim 9$~mag in the N band
($\sim 10$~mJy). MIDI is equipped with narrow band filters, a prism
and a grism, providing maximum spectral resolution of $\sim 530$. In
the last few years the instrument has been used extensively to measure
the distribution of circumstellar dust around AGB and post-AGB
stars. Among the highlights of these observations, the distribution of
silicate dust around a carbon star \citep{Ohnaka2008} and the detection
of circumbinary disks in AGB and post-AGB stars \citep{Deroo2007a,
  Deroo2007b, Deroo2006}

The two Keck telescopes are also capable of working as an interferometer
at infrared wavelength \citep{Colavita2003}. With a baseline of $\sim
85$~m it provides an angular resolution of $\sim 27$~mas in the N
band. This system has been recently equipped with a low resolution
spectrometer ($\Delta \lambda / \lambda ~\simeq 35$)
providing wavelength dependent fringe visibilities across the
8--12~\um{} spectral range \citep{Mennesson2005}. The Keck
interferometer can also operate as a nulling interferometer (KNI,
\citealt{Serabyn2004}), in which case the light coming from the two
telescopes is made to interfere destructively, in order to cancel out
the central point source. This technique has been tested successfully
at the MMT telescope to observe the circumstellar environment of
evolved stars \citep{Hinz1998}, and a dedicated instrument is being
built for the Large Binocular Telescope (LBT, \citealt{Hinz2008}). The
KNI is already accepting proposals, while the LBT interferometer is
currently in construction.

The main limitation of long baseline interferometry is the small
number of apertures that can be combined in the current
setups. Because of this the source coverage in Fourier space is
generally not sufficient to generate true images of the observed
targets. The Magdalena Ridge Observatory Interferometer (MROI) in New
Mexico is being built from the ground up to be an imaging
interferometer \citep{Creech-Eakman2008}. The MROI will operate in the
optical and near-IR, in order to access key gas and dust diagnostic
emission. It will initially have 6 1.4~m aperture telescopes (10 when
completed) distributed in a reconfigurable array (like the VLA), to
achieve angular resolution from 0.1 to 100~mas. Thanks to these
multiple baselines, it will be able to efficiently generate true
images (like current radio interferometers), for sources as faint as
14~mag in the H band (1.6~\um). The interferometer is on track to
produce ``first light'' images by 2010.

While the new generation of imaging interferometers is still in
construction, diffraction limited images have been produced at
infrared wavelengths with large aperture telescopes without AO, using
speckle imaging and aperture masking. The angular resolution is not as
high as in the case of long baseline interferometry, but is sufficient
to probe for symmetry and anisotropy in the circumstellar
envelopes. The working principle of speckle interferometry is the
acquisition of short time exposures that are then combined in Fourier
space \citep{Weigelt1977, Lohman1983, Nisenson1988}. Aperture masking
consists in dividing the aperture of a single mirror telescopes in
sub-apertures via a physical non-redundant mask. Image reconstruction
techniques are then used to recover the final high resolution images
\citep{Haniff1987, Tuthill2000}.

Both techniques have been used in the near-IR to obtain images of AGB
and post-AGB circumstellar dusty envelopes with an angular resolution
of $\sim 50$--70~mas. \citet{Weigelt1998, Weigelt2001} and
\citet{Tuthill2000, Tuthill2005}, in particular, followed the
evolution of inhomogeneous and clumpy mass loss processes within the
dust shell of IRC$+$10216, with a spatial resolution of $\sim 5$~AU
(a few stellar radii). These observations provide a unique look into
the time variability of dust condensation in the AGB outflow, showing
extensive departures from spherical symmetry at the level of the dust
formation zone.

The presence of the ``warm'' mask, which is a strong source of thermal
noise, limits the utility of aperture masking in the thermal IR. The
technique of ``segment tilting'' \citep{Monnier2004, Weiner2006b},
solves this issue by re-orienting groups of the segments of a large
aperture telescope like Keck to produce interferograms from which
independent visibility amplitudes and closure phases can be
measured. This technique has been used to obtain 8, 9.9, 10.7 and
12.5~\um{} images of $o$~Cet, resolving the wind accretion from
Mira A to its low mass companion and detecting a possible long
lived disk around Mira B \citep{Ireland2007}.


\section{Jets and Shocked Gas}
\label{sec-xrays}

In a symbiotic system mass transfer from the evolved giant to a more
compact hot star typically proceeds via Bondi-Hoyle capture. Only a
small fraction of the wind is accreted through this process, which can
only be detected because of the UV radiation emitted by the compact
companion. Occasionally, however, this picture changes as the
symbiotic system goes into an outburst which can have an optical
amplitude of several magnitudes. The remnants of these dramatic
events, which involve bipolar outflows with speed in excess of
1~000~km~s$^{-1}$, can be directly imaged by the HST in narrow band
ionized gas filters (see e.g. \citealt{Corradi2001, Sahai2003}), by radio
interferometers via synchrotron radiation or thermal bremsstrahlung of
the shocked gas (see e.g. \citealt{Crocker2001}) and in soft- and
hard-band X-rays (see e.g. \citealt{Karovska2005, Karovska2007,
  Kellogg2007}).

Imaging the outbursts in nearby symbiotic systems, that can be
spatially resolved with much greater details than jets in more distant
sources like YSOs or QSOs, provides important observational tests for
astrophysical jets models. The peculiarity of AGB mass loss
processes and wind accretion in binary systems is also important for
the role played by these phenomena in the creation of asymmetric
post-AGB and planetary nebulae, and ``extrinsic'' AGB stars polluted
by the mass transfer from a companion on the AGB phase.


\section{Imaging the Star}
\label{sec-star}

Given that, ultimately, the AGB mass loss engine is the AGB star
itself, many of the open questions concerning the geometry of the wind
and its relationship with the stellar pulsations can be investigated
by imaging the stellar photosphere and the molecular layers
immediately on top of the photosphere. This has been possible for
quite a long time. The same interferometric techniques described in
Section~\ref{ssec-interferometers}, at optical and near-IR
wavelengths, can resolve the stellar disks with large aperture
telescopes. At 700~nm, for example, a 6~m telescope can achieve an
angular resolution of $\sim 30$~mas, sufficient to resolve the disk of
nearby AGB stars. Narrow band imaging through the whole
electromagnetic spectrum is particularly useful to isolate individual
atomic and molecular layers in the photospheres, measuring changes in
size and asymmetry. It can also investigate surface brightness
variations in the stellar photosphere (including hot and dark spots)
and derive very precise measurements of stellar limb darkening and
effective temperatures, which are important constraints for AGB
dynamic atmosphere modeling.

Michelson interferometers have been used for a long time to measure
the diameter of giant stars (including AGBs, see e.g. the
\citealt{Pease1931} measurement of the $o$~Cet diameter) at optical
wavelengths. This measurements are carried on to determine stellar
diameter variations in the TiO optical bands with the COAST
interferometer \citep{Burns1998, Young2000}. In the near-IR these
measurements are done with the IOTA/FLUOR \citep{vanBelle1996,
  vanBelle1997, vanBelle2002, Perrin1999}, PTI \citep{Thompson2002a,
  Thompson2002b}, VLTI \citep{Boboltz2005, Woodruff2004, Richichi2003}
and Keck \citep{Eisner2007} interferometers. A strong wavelength
dependence in the diameter of these stars was noticed when
simultaneous observations with broad and narrow band filters
\citep{Mennesson2002, Perrin2004, Millan-Gabet2005, Wittkowski2008,
  Eisner2007} became possible. With closure phases (as in the case of
IOTA/FLUOR and VLTI/AMBER), the measure of asymmetry is much more
precise, and even image reconstruction can be attempted (see
e.g. \citealt{Ragland2006, Ragland2008}).

Speckle interferometry and aperture masking have also been used in the
optical to resolve the disk of AGB stars (see
e.g. \citealt{Karovska1991, Weigelt1996, Wilson1992, Tuthill1994,
  Tuthill1995, Tuthill1999, Haniff1995, Bedding1997} and references
therein). As mentioned in Section~\ref{sec-radio}, AGB H$^-$ free-free
photospheres have also been detected in the radio by \citet{Reid2007},
opening another venue to measure stellar asymmetries.

It is finally worth noting that interferometric techniques have also
been used in space to measure the diameter and symmetry of AGB optical
photospheres. The Fine Guidance Sensor (FGS) on-board the Hubble Space
Telescope (HST) was used to derive the optical angular diameter of AGB
stars with interferometric techniques, finding strong evidences of
asymmetry for R~Leo and W~Hya \citep{Lattanzi1997}. For the nearest
stars, in fact, the stability of the PSF of the HST is sufficient to
resolve the stellar photosphere with direct optical and UV imaging
(see e.g. the image of $o$~Cet obtained by \citealt{Karovska1997}).


\section{The Future}
\label{sec-future}

From the brief review presented in these pages it is clear that
meaningful studies of AGB mass loss can be greatly improved by
simultaneous multi-wavelength observations. Due to their Long Period
Variability, AGB stars change their radius by a
significant fraction on time-scales of days. Modeling data sets
collected at different epochs is complicated by these changes: to
match spectral observations with high resolution imaging the
two datasets must be obtained at nearly the same time (see
e.g. \citealt{Wittkowski2007, Karovikova2008}).

New imaging facilities will become available within the next decade,
opening new opportunities to map AGB circumstellar envelopes from the
ground and from space. ALMA and MROI will provide high sensitivity and
angular resolution in the infrared and sub-mm. The BLT and the KNI
will make nulling interferometry of circumstellar dust practical. A
new generation of 20-30~m segmented telescopes equipped with AO will
increase the sensitivity and resolution achievable with direct
imaging. These breakthroughs in ground based astronomy will be matched
by new space instrumentation. The MIRI instrument on-board the James
Webb Space Telescope (JWST) will provide diffraction limited
narrow band imaging between 5 to 27~\um, sampled on a pixel size of
$\sim 0.11$~arcsec. The NIRCam/JWST camera will instead operate
between 0.6 to 5~\um, providing Nyquist sampled images with resolution
of 0.06--0.13~arcsec. Both instruments will be equipped with
cononographs, a feature that is essential to suppress the bright
central AGB star and detect the faint circumstellar emission. These
observations will be complemented at sub-mm wavelengths by the imaging
photometers of the Herschel Space Telescope, that will have limited
angular resolution but unprecedented sensitivity to detect cold
detached shells.

The development of high angular resolution imaging techniques like
interferometry and adaptive optics, at wavelengths ranging from the
X-rays to the radio, has opened in the last decade the possibility to
study with great detail the physics of mass loss in AGB and post-AGB
stars. The complex view that is arising from these new observations is
certainly a challenge for stellar modelers, but also an opportunity to
finally understand one of the most important, but also neglected,
ingredients in stellar and galactic evolution.




\section*{Acknowledgments} 

I would like to thanks the organizers of the workshop ``The Origin of
the Elements Heavier than Iron'' for the wonderful time spent in
Torino, and Roberto Gallino, to whom the workshop was dedicated, and
the whole stellar group in Torino for their support while I was taking
my first steps in astrophysics.


\end{document}